\input harvmac
\overfullrule=0pt

%
\def\sqr#1#2{{\vbox{\hrule height.#2pt\hbox{\vrule width
.#2pt height#1pt \kern#1pt\vrule width.#2pt}\hrule height.#2pt}}}

\def\half{{\textstyle{1\over 2}}}

\Title{ \vbox{\baselineskip12pt
\hbox{hep-th/0308089}}}
{\vbox{\centerline{A Relation Between Approaches To Integrability}
\bigskip
\centerline{in Superconformal Yang-Mills Theory}}}
\smallskip
\centerline{Louise Dolan
}
\smallskip
\centerline{\it Department of Physics}
\centerline{\it
University of North Carolina, Chapel Hill, NC 27599-3255}
\bigskip
\smallskip
\centerline{Chiara R. Nappi
}
\smallskip
\centerline{\it Department of Physics, Jadwin Hall}
\centerline{\it Princeton University, Princeton, NJ 08540}
\bigskip
\smallskip
\centerline{Edward Witten}
\smallskip
\centerline{\it School of Natural Sciences, Institute for Advanced Study}
\centerline{\it Olden Lane, Princeton, NJ 08540, USA}
\bigskip

\noindent We make contact between the infinite-dimensional
non-local symmetry of the type IIB superstring on $AdS_5\times
S^5$ and a non-abelian infinite-dimensional symmetry algebra for
the weakly coupled superconformal gauge theory. We explain why the
planar limit of the one-loop dilatation operator is the
Hamiltonian of a spin chain, and show that it commutes with the
$g^2 N = 0$ limit of the non-abelian charges.

\Date{}

\nref\Polyakovone{A. M. Polyakov, ``Interaction of Goldstone
Particles in Two Dimensions. Applications to Ferromagnets and
Massive Yang-Mills Fields,'' Phys. Lett. {\bf B 59}, 79 (1975);
``Hidden Symmetry Of The Two-Dimensional Chiral Fields,''
Phys. Lett {\bf B72}, 224 (1977);
``String Representations and Hidden
Symmetries for Gauge Fields,'' Phys. Lett B82 (247) 1979;
``Gauge Fields as Rings of Glue,'' Nucl. Phys. {\bf B164}, 1971 (1980).}

\nref\Dolanone{L.
Dolan, ``Kac-Moody Algebra is Hidden Symmetry of Chiral Models,''
\ Phys.\ Rev. \ Lett. {\bf 47} 1371 (1981);
``Kac-Moody Symmetry Group of Real Self-dual
Yang-Mills,'' \ Phys. \ Lett. {\bf 113B}, 378 (1982);
``Kac-Moody Algebras and Exact
Solvability in Hadronic Physics'', \ Phys. \ Rep. {\bf 109}, 1
(1984).}

\nref\wu{L. Chau, M. Ge, and Y.S. Wu, ``The Kac-Moody Algebra in the
Selfdual Yang-Mills Equation'', Phys.\ Rev.\ D {\bf 25}, 1086 (1982).}

\nref\un{K.Ueno and Y. Nakamura,
``Transformation Theory For Anti(Self)Dual Equations And
The Riemann-Hilbert Problem,''
Phys.\ Lett.\ B {\bf 109} (1982) 273.}

\nref\pop{A.~D.~Popov and C.~R.~Preitschopf,
``Conformal Symmetries of the Self-Dual Yang-Mills Equations,''
Phys.\ Lett.\ B {\bf 374}, 71 (1996)
arXiv:hep-th/9512130.}

\nref\thooft{G. 't Hooft, ``A Planar Diagram Theory for Strong
Interactions,'' Nucl. Phys. {\bf B72}, 461 (1974).}

\nref\lipatov{L. N. Lipatov, ``High-energy Asymptotics of Multicolor
QCD and Exactly Solvable Lattice Models,'' JETP Lett. {\bf 59}, 596 (1994),
arXiv:hep-th/9311037.}

\nref\Korchemsky{G. Korchemsky and L. Faddeev,
`High-energy QCD as a Completely Integrable Model,''
Phys.\ Lett.\ B {\bf 342} 311 (1995),
arXiv:hep-th/9404173.}

\nref\bmn{D. Berenstein, J. Maldacena, and H. Nastase,
``Strings in Flat Space and pp Waves from $N=4$ Super Yang Mills,''
JHEP {\bf 204}, 013 (2002).
arXiv:hep-th/0202021.}

\nref\Minahan{J.~A.~Minahan and K.~Zarembo, ``The Bethe-ansatz for
N = 4 super Yang-Mills,'' JHEP {\bf 0303}, 013 (2003)
arXiv:hep-th/0212208.}

\nref\Beisertone{N. Beisert, ``The Complete One-loop Dilatation
operator of N = 4 super Yang-Mills Theory,''
arXiv:hep-th/0307015.}

\nref\Beiserttwo{N.~Beisert and M.~Staudacher, ``The N = 4 SYM
Integrable Super Spin Chain,'' arXiv:hep-th/0307042.}

\nref\Beisertthree{N.~Beisert, J.~A.~Minahan, M.~Staudacher and
K.~Zarembo, ``Stringing Spins and Spinning Strings,''
arXiv:hep-th/0306139.}

\nref\Beisertfour{N.~Beisert, C.~Kristjansen and M.~Staudacher,
``The Dilatation Operator of N = 4 Super Yang-Mills theory,''
Nucl.\ Phys.\ {\bf B664}, 131 (2003) arXiv:hep-th/0303060.}

\nref\Belitskyone{A. ~Belitsky, A. ~Gorsky and G. ~Korchemsky,
``Gauge / String Duality for QCD Conformal Operators,'' hep-th/0304028.}

\nref\Kotikov{A.~V.~Kotikov, L.~N.~Lipatov and V.~N.~Velizhanin,
``Anomalous Dimensions of Wilson operators in N = 4 SYM theory,''
Phys.\ Lett.\ B {\bf 557} 114 (2003),
arXiv:hep-ph/0301021.}

\nref\Osborn{F.~A.~Dolan and H.~Osborn,
``Superconformal S1ymmetry, Correlation Functions and the Operator Product  
Expansion,''
Nucl.\ Phys.\ B {\bf 629}, 3 (2002),
arXiv:hep-th/0112251.}

\nref\Braun{V.~M.~Braun, S.~E.~Derkachov and A.~N.~Manashov,
``Integrability of Three-particle Evolution Equations in {QCD},''
Phys.\ Rev.\ Lett.\  {\bf 81}  2020 (1998).
arXiv:hep-ph/9805225.}

\nref\Belitskytwo{A.~V.~Belitsky,
``Fine Structure of Spectrum of Twist-three Operators in {QCD},''
Phys.\ Lett.\ B {\bf 453} 59 (1999),
arXiv:hep-ph/9902361; 
``Integrability and WKB solution of Twist-three Evolution Equations,''
Nucl.\ Phys.\ B {\bf 558} 259 (1999),
arXiv:hep-ph/9903512;
`Renormalization of Twist-three Operators and Integrable Lattice Models,''
Nucl.\ Phys.\ B {\bf 574} 407 (2000),
arXiv:hep-ph/9907420.}

\nref\Manashov{V.~M.~Braun, S.~E.~Derkachov, G.~P.~Korchemsky 
and A.~N.~Manashov,
``Baryon Distribution Amplitudes in {QCD},''
Nucl.\ Phys.\ B {\bf 553} 355 (1999),
arXiv:hep-ph/9902375.}

\nref\Derkachov{S.~E.~Derkachov, G.~P.~Korchemsky and A.~N.~Manashov,
``Evolution Equations for Quark Gluon Distributions in Multi-color QCD  
and Open Spin Chains,''
Nucl.\ Phys.\ B {\bf 566} 203 (2000),
arXiv:hep-ph/9909539.}

\nref\wadia{G. Mandal, N. V. Suryanarayana, and S.R. Wadia,
``Aspects of Semiclassical Strings in AdS(5),'' Phys. Lett. {\bf B543},
81 (2002), arXiv:hep-th/0206103.}

\nref\Metsaev1{R.~R.~Metsaev and A.~A.~Tseytlin, ``Type IIB
Superstring Action in AdS(5) x S(5) Background,'' Nucl.\ Phys.\
{\bf B533}, 109 (1998) arXiv:hep-th/9805028.}

\nref\Bena{I. Bena, J. Polchinski and R. Roiban, ``Hidden
Symmetries of the AdS(5) x $S^5$ Superstring,''
arXiv:hep-th/0305116.}

\nref\fhj{S. Fubini, A. Hanson, and R. Jackiw,
``New Appproach to Field Theory'', Phys. Rev. {\bf D7}, 1732 (1973).}

\nref\LP{M. Luscher and K. Pohlmeyer, ``Scattering of Massless
Lumps and Nonlocal Charges in the Two-dimensional Classical
Non-linear Sigma Model'',  Nucl. Phys. {\bf B137}, 46 (1978).}

\nref\Luscher{M. Luscher, ``Quantum Non-local Charges and Absence of Particle
Production in the Two-dimensional Non-linear $\sigma$ Model'',
\ Nucl.\  Phys. {\bf B135}, 1 (1978).}

\nref\faddeev{ L. A. Takhtadjan and L. Fadde'ev, {\it Hamiltonian
Methods in the Theory of Solitons}, Springer Series in Soviet
Mathematics, Springer Verlag, 1987; L. Fadde'ev, ``How Algebraic Bethe Ansatz
works for Integrable Models'', hep-th/9605187.}

\nref\berkstu{B.~C.~Vallilo, ``Flat Currents in the Classical
${\rm AdS}_5 \times  S^5$ Pure Spinor Superstring,''
arXiv:hep-th/0307018.}

\nref\berk{N.~Berkovits,
``Super-Poincare Covariant Quantization of the Superstring,''
JHEP {\bf 0004}, 018 (2000),
arXiv:hep-th/0001035.}

\nref\schwarz{J.~H.~Schwarz,
``Classical Symmetries of Some Two-dimensional Models Coupled to Gravity,''
Nucl.\ Phys.\ B {\bf 454}, 427 (1995)
arXiv:hep-th/9506076.}

\nref\GO{P. .~Goddard and D.~I.~Olive,
``Kac-Moody And Virasoro Algebras: A Reprint Volume For Physicists,''
Adv.\ Ser.\ Math.\ Phys.\  {\bf 3} (1988) 1.}

\nref\Bernardtwo{D. Bernard,
``An Introduction to Yangian Symmetries,''
\ Int. \ J.\ Mod.\ Phys.  {\bf B7}, 3517 (1993)
arXiv:hep-th/9211133.}

\nref\mackay{N. J.~MacKay,
``On the Classical Origins of Yangian Symmetry in Integrable Field Theory,''
Phys.\ Lett.\ B {\bf 281}, 90 (1992).}

\nref\LC{D.~Bernard and A.~Leclair,
``Quantum Group Symmetries And Nonlocal Currents In 2-D Qft,''
Commun.\ Math.\ Phys.\  {\bf 142} (1991) 99.}

\nref\Bernardone{D. Bernard,
``Hidden Yangian in 2D Massive Current Algebras,''
Commun.Math.Phys. {\bf 137}, 191 (1991).}

\nref\sw{N.~Seiberg and E.~Witten,
``String Theory and Noncommutative Geometry,''
JHEP {\bf 9909}, 032 (1999)
arXiv:hep-th/9908142.}

\nref\Mikhailov{A. Mikhailov, ``Notes on Higher Spin Symmetries,''
arXiv:hep-th/0201019.}


\newsec{Introduction}

It has long been conjectured \refs{\Polyakovone }
that there might
be integrable structures in four-dimensional quantum gauge theory,
analogous to the known integrable structures in two-dimensional
sigma models and possibly extending what is known for self-dual
gauge fields in four dimensions (see for example \refs{\Dolanone -
\pop}). Any such result is bound to be related to the planar or
large $N$ limit of gauge theories \thooft, for a simple reason.
There is no chance that quantum $SU(N)$ gauge theory would turn
out to be integrable for any given $N$, because the phenomena it
describes (such as nuclear physics for $N=3$ when quarks are
included) are far too complicated. But the phenomena are believed
to simplify for $N\to \infty$ (for example, confining theories
become free in this limit), making integrability conceivable.

A possible clue of this has appeared some time ago in studies of
high energy scattering in gauge theories \refs{\lipatov, \Korchemsky}.
Lately, two
different developments have pointed to integrable structures in
the large $N$ limit of four-dimensional ${\cal N}=4$
supersymmetric Yang-Mills theory (SYM).  One development, which was
stimulated by the BMN model of the plane wave limit of ${\rm
AdS}_5\times S^5$ \bmn, has involved the study of the dilatation
operator in perturbation theory.  Initially in some special cases
\Minahan\ and later in greater generality
\refs{\Beisertone,\Beiserttwo} following additional work
\refs{\Beisertthree,\Beisertfour,\Belitskyone}, it has been argued that the
one-loop anomalous dimension operator can be interpreted as one of
the commuting Hamiltonians of an integrable spin chain.
This result depends on the one-loop anomalous dimensions of twist two
operators, which were also computed
in \refs{\Kotikov,\Osborn}.
That the one-loop anomalous dimension operator  in non-supersymmetric gauge
theory is integrable asymptotically for large angular momentum at least for a
very large class of operators had been
discovered earlier in a parallel line of development
\refs{\Braun, \Belitskytwo, \Manashov, \Derkachov }.

The other development, for which the bosonic theory gave a clue
\wadia, is that the  classical Green-Schwarz superstring action
for ${\rm AdS}_5\times S^5$,  constructed in \Metsaev, has turned
out \Bena\ to possess a hierarchy of non-local symmetries,
presumably implying that the world-sheet theory is a
two-dimensional integrable system, analogous to many other such
systems that are known. (See \refs{\LP, \Luscher}  for
construction of nonlocal conserved charges in sigma models, and
\faddeev\ for an extensive introduction to a variety of types of
integrable two-dimensional model.) Experience with other
two-dimensional systems indicates that the non-local nature of
these symmetries gives them the potential to constrain the
perturbative string spectrum (i.e. the gauge theory spectrum at
$N=\infty$) without usefully constraining the exact string
amplitudes summed over genus (corresponding to an exact solution
of gauge theory at all $N$ -- too much to ask for, as noted
above). One does, however, hope that these symmetries are relevant
to gauge theory in the large $N$ limit for all values of $g^2N$,
not just at $g^2N=\infty$ where the classical analysis in \Bena\
applies. A step in this direction has been obtained by showing
\berkstu\ that analogous non-local symmetries hold in the
Berkovits description \berk\ of ${\rm AdS}_5\times S^5 $.

The present paper aims at a modest step toward relating these two
developments.  Starting with the non-local symmetries of \Bena, we
will try to deduce {\it why} the one-loop anomalous dimension
operator is the Hamiltonian of an integrable spin chain. We begin
in section 2 by guessing how the non-local symmetries should act
on a chain of Yang-Mills partons at $g^2N=0$.  The symmetries
generate a non-abelian algebra that has been called the Yangian,
and there is a natural (and standard) way for the Yangian to be
realized in a chain of partons or spins.

We conjecture that this is the $g^2N=0$ limit of how the
Yangian operators act in Yang-Mills theory. We then argue, also in
section 2, that the one-loop anomalous dimension operator must
commute with the $g^2N=0$ limit of the Yangian. Then in the rest
of the paper, we verify the commutativity explicitly using
formulas and properties developed in
\refs{\Beisertone,\Beiserttwo}. Generally, for $1+1$-dimensional
systems, the local operators that commute with the Yangian are the
integrable Hamiltonians, so this commutativity means that the
one-loop dilatation operator is the Hamiltonian of an integrable
spin chain. See for example \Bernardtwo\ .

The argument showing that the one-loop anomalous dimension
operator commutes with the $g^2N=0$ limit of the Yangian is
special to one loop.  Beyond one loop, we do not expect the
dilatation operator to commute with the Yangian.  The general
structure is that, like all the generators of the superconformal
group $PSU(2,2|4)$, the exact dilatation operator is one of the
generators of the Yangian. Many of the generators of the Yangian,
including the dilatation operator, receive perturbative
corrections beyond one loop.  For example, in higher orders, there
are perturbative corrections to the dilatation operator that do
not conserve the number of partons  (i.e. the length $L$ of the
spin chain), so in general this system cannot be viewed as a
conventional spin chain with the partons as spins.  Certainly,
then, in general the Yangian generators receive corrections.

For some special sets of states, such as sets considered in
\refs{\bmn,\Minahan,\Beiserttwo}, the quantum numbers are such as
to prevent creation and annihilation of partons, and it is
plausible  (and has been proposed) that in such such sets of
states, the exact dilatation operator is the Hamiltonian of an
integrable spin chain. It may be that the higher generators $Q^A$
of the Yangian have no corrections when restricted to such
sectors; this might lead to an interpretation of the exact
dilatation operator as an integrable Hamiltonian in such a sector.
Of the appendices in this paper, only Appendix C
develops material that is actually used in the main text.

\newsec{Non-Local Generators}

We begin by recalling how non-local symmetries arise in
two-dimensional sigma models \LP.  One considers a model with a
group $G$ of symmetries; the Lie algebra of $G$ has generators
$T_A$ obeying $[T_A,T_B]=f_{AB}^CT_C$.  The action of $G$ is
generated by a current $j^{\mu\,A}$ that is conserved,
$\partial_\mu j^{\mu\, A} =0$.  Nonlocal charges arise if in
addition the Lie algebra valued current $j_\mu = \sum_A j_\mu^ A
T_A$ can be interpreted as a flat connection,
\eqn\flatco{\partial_\mu j_\nu-\partial_\nu j_\mu + [j_\mu ,
j_\nu]=0.} (Indices of $j_\mu$ are raised and lowered using the
Lorentz metric in two dimensions.) The conservation of $j_\mu$
leads in the usual fashion to the existence of conserved charges
that generate the action of $G$:
\eqn\hobbo{J^A=\int_{-\infty}^\infty dx \,j^{0\,A}(x,t).} In
addition, a short computation using \flatco\ reveals that
\eqn\jobbo{Q^A = f^A_{BC} \int_{-\infty}^\infty dx\,\int_x^\infty
dy\, j^{0\,B}(x,t) \, j^{0\,C}(y,t) - 2 \int_{-\infty}^\infty dx
\, j_1^A(x,t) } is also conserved.  (For the moment we take the
spatial direction to be noncompact, although in string theory it
is more relevant to compactify on a circle with periodic boundary
conditions.  When one does compactify, $Q^A$ cannot be defined, as
the restriction to $x<y$ does not make sense.) Under repeated
commutators, the $Q^A$ generates an infinite-dimensional symmetry
algebra that has been called the Yangian.   The Yangian has a
basis ${\cal J}_n^A$ where ${\cal J}_0^A=J^A$, ${\cal J}_1^A=Q^A$,
and ${\cal J}_n^A$ is an $n$-local operator that arises in the
$(n-1)$-form commutator of $Q$'s.  Since we will work in this
paper mainly with the generators $J^A$ and $Q^A$, we have given
them those special names.

The detailed algebraic structure of the Yangian is rather
complicated and will not be needed in the present paper.  We
pause, however, to briefly explain how the Yangian is related to
the much simpler partial Kac-Moody algebra that can also be
defined in such systems \refs{\Dolanone, \schwarz,
\GO}. (This is also explained briefly in \refs{\Bernardtwo -
\Bernardone}.)  The Yangian generates by Poisson brackets some
transformations of the fields $\Phi$ that we write schematically
$\delta\Phi = \sum_{n,A}\epsilon_{n,A}{\cal J}_n^A(\Phi)$, where
$\epsilon_{n,A}$ are infinitesimal parameters.   These
transformations are symmetries of the classical equations of
motion, since the Yangian generators commute with the Hamiltonian.
The partial Kac-Moody algebra is generated by infinitesimal
transformations
$\delta\Phi=\sum_{n,A}\tilde\epsilon_{n,A}\delta_{n}^{A}(\Phi)$,
where $\tilde\epsilon_{n,A}$ are another set of parameters and the
objects $\delta_{n}^{A}$ are certain infinitesimal symmetries of
the equations.  Usually, one considers transformations with
field-independent coefficients $\epsilon$ or $\tilde\epsilon$.
However, the equivalence relation generated by the symmetries is
obtained by letting $\epsilon$ and $\tilde\epsilon$ be arbitrary,
so it does not matter if they are field-dependent. In the problem
at hand, the Yangian and partial Kac-Moody symmetries are
different, but become equivalent if one lets $\epsilon$ (or
$\tilde\epsilon$) be field-dependent. This is shown explicitly in
Appendix A.  Thus the Yangian and partial Kac-Moody algebras are
different but generate the same equivalence relation. The relation
between them is somewhat similar to the relation between
commutative and non-commutative Yang-Mills gauge transformations
as given by the Seiberg-Witten map \sw.

There are also discrete spin systems, that is systems in which the
dynamical variables live on a one-dimensional lattice rather than
on the real line, that have the same Yangian symmetry.  The
lattice definition of $J^A$ is clear.  We assume that the spins at
each site $i$ have $G$ symmetry, and let $J^A_i$ be the symmetry
operators at the $i^{th}$ site.  The total charge generator for
the whole system is then \eqn\uncu{J^A=\sum_i J^A_i.} What about
$Q^A$?  At least the bilocal part of \jobbo\ has an obvious
discretization: \eqn\nobbo{Q^A=f^A_{BC}\sum_{i<j}J^B_iJ^C_j.} This
turns out to be the right formula, in many of the most commonly
studied lattice integrable systems.  Note that one has made no
attempt to discretize the second term in \jobbo.  This proves to
be unnecessary.\foot{In many simple models, it is also impossible
for elementary reasons.  One would expect a discretization of
$\int dx \,j_1^A$ to be of the form $\sum_i j_i^A$ where $j_i^A$
acts on the $i^{th}$ site  and transforms in the adjoint
representation. Frequently, there is no such operator except
$J_i^A$.  But taking $j_i^A$ to be a multiple of $J_i^A$ would
simply add to $Q^A$ a multiple of $J^A$; this is a change of basis
that does not affect the structure of the algebra.}  The bare
generators \uncu\ and \nobbo\ satisfy: \eqn\begyan{\eqalign{[J^A,
J^B]&= f_{ABC} J^C\cr [J^A, Q^B] &= f_{ABC} Q^C.\cr}}  An analog
of \nobbo\ in gauge theory at $g^2N=0$ is described in Appendix B.

One generator of $PSU(2,2|4)$, called the dilatation generator
$D$, is of special importance.
In the radial quantization of four-dimensional superconformal
Yang Mills theory on ${\bf R}\times S^3$, $D$ is the Hamiltonian.
\foot{ $D$ is conjugate in  $PSU(2,2|4)$ to $\half (P_0 + K_0)$,
where $P_\mu$, $K_\mu$ are the translations and special conformal
transformations. For radial quantization on a hyperboloid, see
\fhj\ .} Using conformal invariance to identify ${\bf
R}\times S^3$ with ${\bf R}^4$, the states of the quantum theory
on ${\bf R}\times S^3$
 are in one-to-one correspondence with local operators ${\cal O}
(x)$, where the correspondence is $|{\cal O}\rangle \sim
\lim_{x\rightarrow 0} {\cal O}(x) |0\rangle$. In the large $N$
limit of the gauge theory, we focus on single-trace local
operators. Such an operator is the trace of a  product of  letters
where a letter is as follows.   A letter is one of the elementary
fields $\phi^I = \phi^{I{\cal A}}(x) {\cal T}^{\cal A}$,
$\psi_\alpha^a = \psi_\alpha^{a {\cal A}}(x) {\cal T}^{\cal A}$,
$F_{\mu\nu}= F_{\mu\nu}^{\cal A}(x) {\cal T}^{\cal A}$, or the
(symmetrized) $n^{th}$ derivative of one of those, for some $n>0$.
(The indices $1\le I\le 6$, $1\le a\le 4$ label the vector and
spinor $SU(4)$ $R$-symmetry representations.) A single-trace
operator ${\cal O}(x)$ is said to be of length $L$ if it is the trace
of a product of $L$ letters. In the correspondence between
operators and states, the letters form a basis for the
one-particle states of the free ${\cal N}=4$ vector multiplet on
${\bf R}\times S^3$.

We really want to consider a  gauge-invariant state that is a
single trace ${\cal O}(x)=\Tr \,\Phi_1(x)\Phi_2(x)\dots \Phi_L(x)$
of a possibly very large number of fields $\Phi_i$, each of which
is one of the letters considered above. As in many papers cited in
the introduction, we think of the choice of a given ${\cal O}$ as
representing in free field theory a state of a chain of $L$
``spins'' (which we also call ``partons'').   Our ``spins,''
therefore, are simply one-particle states in the ${\cal N} =4$
vector multiplet quantized on $S^3$. In identifying the possible
operators ${\cal O}$ with the states of a spin chain, one ignores
the cyclic symmetry of the trace. One studies all possible states
of the spin chain, even though in the application to gauge theory
one only wants the (gauge invariant) cyclically symmetric states.

Our basic assumption in this paper is that in ${\cal N} =4$ super
Yang-Mills theory at $g^2N=0$, with $J_i^A$ understood as the
$PSU(2,2|4)$ generators of the $i^{th}$ parton, \nobbo\ is the
correct formula for the Yangian generators $Q^A$. Our assumption,
in other words, is that the bilocal symmetry deduced from \Bena\
goes over to \nobbo\ for $g^2N\to 0$. Of course, in any case
\uncu\ is the appropriate free field formula for the $J^A$, so we
do not need to state any hypothesis for these generators. And no
further assumption is needed for the higher charges in the
Yangian; they are generated by repeated commutators of the $Q^A$.
So our hypothesis about $Q^A$ completely determines the form of
the Yangian in the free-field limit.

Now we consider what happens when $g^2N$ is not quite zero.  Some
generators of the Yangian do not receive quantum corrections.  For
example, the spatial translation symmetries and the Lorentz
generators are uncorrected, because the theory can be regularized
in a way that preserves them. But the dilatation operator $D$ --
the generator of scale transformations -- certainly is corrected.
The corrections to the eigenvalues of $D$ are called anomalous
dimensions.

We assume, in view of \Bena, that the ${\cal N}=4$ Yang-Mills
theory in the planar limit does have Yangian symmetry for all
$g^2N$. If so, the corrections modify the form of the generators,
but preserve the commutation relations.  One of the commutation
relations says that $Q^A$ transforms in the adjoint representation
of the global group $PSU(2,2|4)$ generated by $J^A$:
$[J^A,Q^B]=f^{ABC}Q^C$. We will write $J^A$ and $Q^A$ for the
charges at $g^2N=0$, and $\delta J^A$ and $\delta Q^A$ for the
corrections to them of order $g^2N$. We write ${\tilde J}^A$ and
${\tilde Q}^A$ for the exact generators (which depend on $g^2N$), so
${\tilde J}^A=J^A+(g^2N)\delta J^A+{\cal O}((g^2N)^2)$, and likewise
for ${\tilde Q}^A$. To preserve the commutation relations, we have
\eqn\incoco{[\delta J^A,Q^B]+[J^A,\delta Q^B]=f^{ABC}\delta Q^C.}
We are now going to make an argument for the Yangian that
parallels one used in \Beisertone\ for the $PSU(2,2|4)$
generators. We consider the special case of this relation in which
$A$ is chosen so that $J^A$ is the dilatation operator $D$. We
also pick a basis  $Q^B$ of the $Q$'s to diagonalize the action of
$D$, so the $PSU(2,2|4)$ algebra reads in part
$[D,Q^B]=\lambda^BQ^B$, where $\lambda^B$ is the bare conformal
dimension of $Q^B$.  Then \incoco\ gives us \eqn\pico{[\delta
D,Q^B]+[D,\delta Q^B] = \lambda^B \delta Q^B.} However, in
perturbation theory, operators only mix with other operators of
the same classical dimension.  So just as $[D,Q^B]=\lambda^B Q^B$,
we have $[D,\delta Q^B]=\lambda^B \delta Q^B$. Combining this with
\incoco, we have therefore \eqn\tuggo{[\delta D,Q^B]=0.} Precisely
the same argument was used in \Beisertone\ to show that $[\delta
D,J^A]=0$; this was a step in determining $\delta D$. Combining
this with \tuggo, we see that $\delta D$ must commute with the
$g^2N=0$ limit of the whole Yangian.

The structure of perturbation theory implies in addition that the
operator $\delta D$ is a sum of operators local along the chain;
this fact has been exploited in \bmn\ and many subsequent papers.
(In fact, $\delta D$, as described explicitly in \Beisertone, is a
sum of operators that act on nearest neighbor pairs.)  The
operators of this type that commute with the Yangian -- where here
we mean the Yangian representation most commonly studied in
lattice integrable models, which for us is the one generated at
$g^2N=0$ by $J^A$ and $Q^A$ -- are called the Hamiltonians of the
integrable spin chain. Thus, from our assumption about the
free-field limit of the Yangian, we have been able,  starting with
the nonlocal symmetries found in \Bena, to deduce the basic
conclusion of \Beiserttwo, found earlier in a special case in
\Minahan,  that $\delta D$ is a Hamiltonian of an integrable spin
chain.

In the remainder of this paper, we will verify this picture by
proving directly, using formulas from
\refs{\Beisertone,\Beiserttwo}, that it is true that $\delta D$
commutes with the Yangian.  Since its commutativity with $J^A$ was
already used in \Beisertone\ to compute $\delta D$, we only need
to verify that $[\delta D, Q^A]=0$.

From what we have said, it is clear that the appearance of a
Hamiltonian that commutes with the Yangian depends on expanding to
first order near $g^2N=0$.  In the exact theory, at a nonzero
value of $g^2N$, one would simply say that the exact dilatation
operator ${\cal D}$, which of course depends on $g^2N$, is one of
the generators of the Yangian. It is not the case in the exact
theory that one has a Yangian algebra and also a dilatation
operator that commutes with it.

In string theory, or Yang-Mills theory, one really wants to
compactify the string (or the spin chain) on a circle with
periodic boundary conditions, since the string is closed. This
makes it impossible to define the Yangian, because the restriction
of the integration region in \jobbo\ to $x<y$ does not make sense.
The global $PSU(2,2|4)$ generators $J^A$ still make sense, of
course, and so do some globally defined operators -- traces of
holonomies, which one might think of as Casimir operators of the
Yangian.  These Casimir operators, which commute with
$PSU(2,2|4)$, perhaps can be used as an aid in computing the
spectrum of ${\cal N}=4$ super Yang-Mills theory in the planar
limit.
Some of these Casimir operators are odd under charge conjugation
(which is the symmetry that
reverses the order of the spin chain), so the fact that they commute with
$PSU(2,2|4)$ would lead
to degeneracies among states of opposite charge conjugation properties, as
found and exploited in
\Beisertfour.

\newsec{Commutation of $Q^A$ with the Planar One-Loop Hamiltonian}

Now we will prove that \tuggo\ \eqn\tuggoone{[\delta D, Q^A ] = 0}
is true, using the properties of $\delta D$ for the super Yang
Mills theory. In this section, to simplify notation, and in view
of its interpretation as the Hamiltonian of an integrable spin
chain, we refer to $\delta D$ as $H$. Actually, we will show that
the commutator $[H,Q^A]$ is the lattice version of a total
derivative, in the following sense. The general form of $H$ for a
chain of length $L$ is that it is a sum of operators each of which
only acts on nearest neighbors, \eqn\ham{H = \sum_{i=1}^{L-1}
H_{i, i+1}\,.} A lattice version of a total derivative is an
expression such as
 \eqn\derdif{q^A = \sum_{i=1}^{L-1} \left(J^A_i -
J^A_{i+1}\right) = J^A_1 - J^A_{\rm L},} which is a sum of
difference operators along the chain and only acts at the ends of
the chain.

We will show, using the specific form of $H$ determined in
\Beisertone, that \eqn\qhcom{[H,Q^A]=q^A,} where $q^A$ is such a
total derivative.  For an infinite chain (which would correspond
more closely to the $1+1$-dimensional field theory studied in
\Bena) and assuming no spontaneous symmetry breaking so that
surface terms at infinity can be dropped, the total derivative
term in \qhcom\ vanishes, and in that sense $[H,Q^A]=0$ for an
infinite chain. For a finite chain with periodic boundary
conditions, the situation is similar though more subtle. \qhcom\
together with the fact that $[H,J^A]=0$ implies that the
commutator of $H$ with any generator of the Yangian is the
integral of a lattice total derivative. Therefore, for a finite
chain with periodic boundary conditions, where a total derivative
will sum to zero, the commutator of $H$ with a Casimir operator of
the Yangian (which is well-defined with periodic boundary
conditions) vanishes.

Let $V_F$ be the space of one-particle states in free ${\cal N} =
4$ super Yang-Mills theory on ${\bf R}\times S^3$.  (Thus the set
of letters as defined above is a vector space basis for $V_F$.) In
the spin chain that is relevant to planar Yang-Mills theory at
$g^2N=0$,  $V_F$ is the space of states of a single spin. $H_{12}$
acts on a two-spin system, which as a representation of
$PSU(2,2|4)$ is simply $V_F\otimes V_F$. The decomposition of
$V_F\otimes V_F$ in irreducible representations of $PSU(2,2|4)$ is
surprisingly simple and plays an important role in \Beisertone.
The decomposition, which we heuristically explain in Appendix C,
is \eqn\tp{V_F \otimes V_F = \bigoplus_{j=0}^\infty V_j,} where,
apart from some exceptions at small $j$, $V_j$ can be
characterized as a representation whose superconformal primary (or
highest weight vector) is an $R$-singlet of angular momentum
$j-2$. ($\bigoplus_{j=0}^\infty V_j $ merely designates a direct
sum of modules $V_j$, and {\it not} any sum on lattice sites.) We
also will need to know how the $PSU(2,2|4)$ quadratic Casimir
operator $J^2 =\sum_A J^A J^A$ (described more precisely in
Appendix C) acts on $V_F\otimes V_F$. With $J^A_1$ and $J^A_2$
denoting the $PSU(2,2|4)$ generators of the first and second
spins, respectively, the quadratic Casimir operator of the
two-spin system is \eqn\qcot{J^2_{12} = \sum_A (J^A_1 + J^A_2)
(J^A_1 + J^A_2).}  Astonishingly, just as if the group were
$SU(2)$ instead of $PSU(2,2|4)$, this operator has eigenvalue $ j
(j+1)$ when acting on $V_j$: \eqn\ev{J^2_{12} V_j = j (j+1) V_j\,,
\quad j= 0,1,2,\dots\,.} This fact is used in \Beisertone\ and
below; we give a brief explanation of it in Appendix C.

According to \Beisertone, $H$ is a sum of two-body operators as in
\ham, where the basic two-body operator is \eqn\engy{H_{12} =
\sum_{j=0}^\infty 2 h(j) P_{12,j}\,.} Here $P_{12,j}$ is the
operator that projects the two-body Hilbert space $V_F\otimes V_F$
onto $V_j$, and $h(j)$ are the harmonic numbers $h(j) =
\sum_{n=1}^j {1\over n}$ for $j\in {{\bf  Z}_+}$ (one defines
$h(0) = 0$).  According to our hypothesis \nobbo, the bilocal
Yangian generator $Q^A$ is also a (non-local) sum of two-body
operators, $Q_A=\sum_{i<j} Q_{ij}^A$, where the basic two-body
operator is \eqn\horel{Q^A_{ij}=\sum_{B,C}f^A_{BC} J^B_iJ^C_j.} We
observe the identity \eqn\anothrel{Q^A_{ij} = {\textstyle{1\over
4}} \,[\, J^2_{ij} , q^A_{ij} \,]\,,} where $J^2_{ij}$ is the
quadratic Casimir operator of the two-particle system, and
$q^A_{ij}$ is the difference operator
\eqn\torel{q^A_{ij}=J^A_i-J^A_j.}

We will first  prove \qhcom\ for a system of two spins.  For this
purpose, we use \anothrel\ to write \eqn\prfone{\eqalign{ [
H_{12}, Q^A_{12} ] &= \textstyle{1\over 4} [ H_{12}, [J^2_{12},
q^A_{12}] ]\,.\cr}}
Then acting on a two-particle state $|\lambda(j)\rangle$ that is
contained in $V_j$ (and so has eigenvalues of $H_{12}$ and
$J^2_{12}$ given above), we have \eqn\prftwo{\eqalign{ [ H_{12},
Q^A_{12} ] |\lambda(j)\rangle&= \textstyle{1\over 4} \big ( H_{12}
J^2_{12} - j(j+1) H_{12} - 2 h(j) J^2_{12} + 2 h(j) j (j+1) \big )
\,\, q^A_{12} |\lambda(j)\rangle\,.\cr}} We will show in section
3.1 that the action of $q_{12}^A$ on a state in $V_j$ can be
written as a linear combination of states in $V_{j-1}$ and
$V_{j+1}$, {\it i.e.} for any $|\lambda(j)\rangle\in V_j$, we have
\eqn\decomp{q_{12}^A |\lambda(j)\rangle = |\chi^A(j-1)\rangle +
|\rho^A (j+1)\rangle,} where $|\chi^A(j-1)\rangle\in V_{j-1}$ and
$|\rho^A(j+1)\rangle\in V_{j+1}$. Given this fact, from \prftwo\
we have \eqn\prfthree{\eqalign{ &[ H_{12}, Q^A_{12} ]
|\lambda(j)\rangle\cr &= j\, (h(j) - h(j-1))  \,\,
|\chi^A(j-1)\rangle \,\, +  \, \,(j+1) (h(j+1) - h(j)) \,\,
|\rho^A (j+1)\rangle \cr &= q^A_{12} |\lambda(j)\rangle\,.\cr}} We
used the fact that $h(j)-h(j-1)=1/j$.

Now let us consider a chain of more than two spins. $H$ is a sum
of nearest neighbor terms $H_{i,i+1}$, while $Q^A$ is a bilocal
sum of two-body operators $Q^A_{j,k}$ with $j<k$.  We have
\eqn\untu{0=\left[H_{i,i+1}\,,\,\sum_{j<k,(j,k)\not= (i,i+1)}
Q^A_{jk}\right].} In fact, terms in the sum in which neither $j$
nor $k$ equals $i$ or $i+1$ are trivially zero. On the other hand,
terms with (say) $k>i+1$ and $j=i,i+1$  add up to
$f^A_{BC}[H_{i,i+1},(J^B_{i}+J^B_{i+1})J^C_k]$, which vanishes
because $J^B_{i}+J^B_{i+1}$ is the $PSU(2,2|4)$ generator of the
two-spin system and so commutes with $H_{i,i+1}$, as does $J^C_k$
for $k>i+1$.

So the commutator $[H,Q^A]$ collapses to
\eqn\prffour{[H,Q^A]=\sum_{i=1}^{L-1}[H_{i,i+1},Q^A_{i,i+1}]
=\sum_{i=1}^{L-1}q^A_{i,i+1}=q^A,} where
$q^A_{i,i+1}=J^A_i-J^A_{i+1}$ is the difference operator of the
two-spin system, and $q^A=J^A_1-J^A_L$.  We have established our
claim that $[H,Q^A]$ is the lattice version of a total derivative.

\subsec{Decomposition of $q^A V_j$ }

In this subsection, we consider a system of two spins, with
$q^A=q^A_{12}=J^A_1-J^A_2$. We wish to show that $q^AV_j$ is
contained in $V_{j+1}\oplus V_{j-1}$.  We do this by proving two
facts:

(1) $q^A V_j$ is contained in the direct sum of $V_k$ with $k-j$
odd.

(2) $q^A V_j$ is contained in the direct sum of $V_k$ with
$|k-j|\leq 1$.

Clearly the two facts together imply what we want.

Fact (1) follows directly by considering the operator $\sigma$
that exchanges the two spins, that is the two copies of $V_F$ in
$V_F\otimes V_F$. (If the two spins are both fermionic, one
exchanges them with a minus sign.) The operator $q^A$ is odd under
$\sigma$.  As explained in Appendix C, $\sigma$ has eigenvalue
$(-1)^j$ on $V_j$.  Fact (1) is a direct consequence of these two
assertions.

To prove fact (2), we first note that to prove that $q^AV_j\subset
\oplus_{k\in T}V_k$, where $T$ is any set of allowed values of $k$
(such as the set $|j-k|\leq 1$) it suffices to show that if
$|\psi(j)\rangle$ is a primary state in $V_j$, then
\eqn\hurglo{q^A|\psi(j)\rangle\in \oplus_{k\in T}V_k.} Indeed any
state in $V_j$ is a linear combination of states $L_1L_2\dots
L_s|\psi(j)\rangle$ where the $L$'s are ``raising'' operators in
$PSU(2,2|4)$ (and are, in fact, either momenta $P_\mu$ or global
supersymmetries $Q_\alpha^a$). To study $q^A L_1L_2\dots
L_s|\psi(j)\rangle$, we try to commute $q^A$ to the right so that
we can use \hurglo. In the process we meet commutators
$[q^A,L_i]$, but these are linear combinations of the $q^B$'s, so,
assuming \hurglo\ has been proved for all choices of $A$, it can
still be used.  The net effect is that \hurglo\ implies that
$q^AL_1L_2\dots L_s|\psi(j)\rangle$ is always contained in
$\oplus_{k\in T}V_k$; in other words, with $T$ being the set
$|k-j|\leq 1$, \hurglo\ implies fact (2).

Since the $q^A$ transform the same way as $J^A$, they have the
same dimensions.   The dimensions of the $q^A$ therefore range
from $1$ to $-1$. The value 1 is achieved only for the components
of $q^A$ that transform like the momentum operators $P^\mu$.

Let us first prove that $q^A|\psi(j)\rangle\subset \oplus_{ k\leq
j+1}V_k$.  For $j\geq 2$, this follows simply by
dimension-counting.  In this range, the primary $|\psi(j)\rangle$
has dimension $j$; it is of course the state of lowest dimension
in $V_j$.  The operator $q^A$ has at most dimension 1, so
$q^A|\psi(j)\rangle$ has dimension at most $j+1$ (and this value
is only achieved if $A$ is such that $q^A$ transforms as one of
the momentum operators). Hence $q^A|\psi(j)\rangle\in\oplus_{k\leq
j+1}V_k$.  To reach the same conclusion for $j=0,1$ takes just a
little more care.  For these values of $j$, $|\psi(j)\rangle$ has
dimension 2.  So $q^A|\psi(j)\rangle$ has dimension at most 3, and
must be contained in $\oplus_{k\leq 3}V_k$.  Given this, fact (1)
above implies further  that $q^A|\psi(1)\rangle\in V_0\oplus V_2$,
which is what we wanted to prove for $j=1$.  For $j=0$, fact (1)
implies that $q^A|\psi(0)\rangle\in V_1\oplus V_3$; we wish to
prove that in fact $q^A|\psi(0)\rangle\in V_1$. This follows from
the $SU(4)_R$ symmetry.  The only state in $V_3$ of dimension no
greater than three is the primary state $|\psi(3)\rangle$, which
is $SU(4)_R$-invariant; but no linear combination of the states
$q^A|\psi(0)\rangle$ has this property.

 Finally, we must prove the
opposite inequality $q^A|\psi(j)\rangle \in \oplus_{k\geq
j-1}V_j$. This is equivalent to saying that $\langle
\chi|q^A|\psi(j)\rangle=0$ if $|\chi\rangle\in V_k$ with $k<j-1$.
We will use the fact that $\langle\chi |q^A|\psi(j) \rangle$ is
the complex conjugate of
$\langle\psi(j)|(q^A)^\dagger|\chi\rangle$. (The adjoint is taken
in radial quantization, so for example the adjoint of the momentum
$P_\mu$ is the special conformal generator $K_\mu$.) But
$(q^A)^\dagger$ is a linear combination of the $q^A$'s, so we can
use the result of the previous paragraph to assert that for
$|\chi\rangle \in V_k$, $ |(q^A)^\dagger|\chi\rangle $ is a sum of
states in $V_m$ with $m\leq k+1$; so this state is orthogonal to
$|\psi(j)\rangle $ if $j>k+1$ or in other words if $k<j-1$. This
completes the proof of fact (2). \vskip20pt 

\appendix {A} {\hskip10pt  Field-dependent Transformations - The
Kac-Moody Loop Algebra and the Yangian}

Here we supply some details of the relation between the partial
Kac-Moody algebra that can be defined in various two-dimensional
integrable models and the Yangian.  For definiteness, we consider
the case of the principal chiral model, which is a two-dimensional
model in which the field $g$ takes values in a Lie group $G$. In
these models, along with the Yangian, it is possible to find
\Dolanone\ non-local transformations that obey the algebra of a
partial Kac-Moody algebra, by which we mean simply the algebra
$[T_{A\,n},T_{B\,m}]=f_{AB}^CT_{C\,m+n}$, $m,n\geq 0$.  We write
$\delta^A_n$ for the symmetry transformations corresponding to the
$T_{A\,n}$. $\delta^A_0$ is the standard global symmetry
generator, and coincides with the transformation generated by the
generator $J^A$ of the Yangian.  Just as the Yangian is generated
by $J^A$ and the first non-trivial generator $Q^A$, and partial
Kac-Moody algebra is generated by $\delta^A_0$ and $\delta^A_1$.

In this model, the infinitesimal transformation generated via
Poisson brackets by $Q^A$, the first non-trivial generator of the
Yangian, can be written in terms of the Kac-Moody generators as
\eqn\tran{ \{ Q^A , g(x,t) \} = -\delta^A_1 g(x,t) +\half f_{ABC}
J^B \, \delta^C_0 (g(x,t)).} Thus, they do not coincide, but they
differ by a term involving the Kac-Moody generator $\delta^A_0$
times $J^B$ (by which we mean simply the function of $g$ and its
derivatives which is the Yangian generator $J^B$).  Thus, $Q^A$
does not generate the same symmetry as $\delta^A_1$.  However,
they differ by a field-dependent multiple of $\delta^A_0$.  Since
the symmetry transformation generated by $\delta^A_0$ is also a
generator of the partial Kac-Moody algebra (or of the Yangian), it
follows that if we want to know if two fields $g(x,t)$ can be
related to each other by a symmetry, it does not matter if the
symmetry group we use is generated by the Kac-Moody algebra or the
Yangian.  Any transformation generated by $J^A$ and $Q^A$, with
some coefficients, can be generated by $\delta^A_0$ and
$\delta^A_1$, with some other coefficients.  (The transformation
from one set of coefficients to the other depends on $g(x,t)$
because $J^B$, which appears in \tran, has such a dependence, so
this transformation does not preserve the Kac-Moody or Yangian
commutation relations.)

\appendix {B} {\hskip2pt Noether Currents}

In order to make contact with conventional Noether current
symmetry analysis, we give the expression for the non-local charge
\nobbo\ in terms of the elementary fields of the super Yang Mills
Lagrangian \eqn\lagr{{\cal L} = {1\over {g_{YM}^2}}\, \Tr \left(
\half F_{\mu\nu} F^{\mu\nu} + D_\mu\phi^I D^\mu\phi^I -\half
[\phi^I, \phi^J ]  [\phi^I, \phi^J ] + {\rm fermions}\right)\,.}
For simplicity, we will only consider $A\in so(2,4)$.

In the classical  theory, the symmetry 
currents for the conformal
group are given in terms of the improved energy-momentum tensor by
\eqn\noeth{j^{A\mu} (x) = \kappa^A_\nu \theta^{\mu\nu}(x)\,,}
where $\kappa^A_\mu$ are the conformal Killing vectors, and
\eqn\niem{\theta^{\mu\nu} = 2 \Tr F^{\mu\rho} F^\nu_{\hskip2pt
\rho} + 2 \Tr D^\mu \phi^I D^\nu \phi^I - g^{\mu\nu} {\cal L}\, 
- {1\over 3} \Tr (D^\mu D^\nu - g^{\mu\nu} D_\rho D^\rho ) \phi^I \phi^I\,
+\,{\rm fermions}\,.} The currents \noeth\ are conserved at any
$g^2N$ using the classical interacting equations of motion.

If we set $g^2N=0$, we note that the {\it untraced} matrix
\eqn\nim{\eqalign{(\theta^{\mu\nu})_i^{\hskip2pt j} &=  F^{\mu\rho}
F^\nu_{\hskip2pt \rho} 
+ F^{\nu\rho} F^\mu_{\hskip2pt \rho} 
+ \partial^\mu \phi^I \partial^\nu \phi^I 
+ \partial^\nu \phi^I \partial^\mu \phi^I  
- g^{\mu\nu} ( \half F_{\rho\sigma} F^{\rho\sigma} 
+ \partial_\mu \phi^I \partial^\mu \phi^I )\cr
&\hskip5pt - {1\over 3} (\partial^\mu \partial^\nu - g^{\mu\nu}
\partial_\rho \partial^\rho ) \phi^I \phi^I 
+ {\rm fermions}\,,\cr}} is also conserved, as is
$\kappa^A_\nu (\theta^{\mu\nu})_i^{\hskip2pt j}$. Here $i,j$ are
the matrix labels of the gauge group generators $(T^{\cal
A})_i^{\hskip2pt j}$. It follows that we can construct non-local
conserved charges by \eqn\nsc{Q^{AB...}_0 = \int_{M} \kappa^A_\nu
(\theta^{0\nu})_i^{\hskip2pt j} \int_{M} \kappa^B_\rho
(\theta^{0\rho})_j^{\hskip2pt k}\ldots\,,} where $M$ is an initial
value surface in spacetime. In free field theory, this acts on a
chain of partons rather as \nobbo\ does, but we have no idea how
to extend the definition to $g^2N\not= 0$.

\appendix {C}{\hskip3pt Primary States In The Two-Particle System}

To determine the one-loop dilatation operator in \Beisertone, it
is necessary to know the decomposition of the two-particle system
in free ${\cal N}=4$ super Yang-Mills theory in irreducible
representations of $PSU(2,2|4)$. This decomposition is
surprisingly simple.  The irreducible representations are
conveniently denoted as $V_j$, $j=0,1,2,\dots$, where the quantum
numbers of a superconformal primary $|\psi(j)\rangle\in V_j$
can be conveniently described as follows.

First of all, the representations $V_0$ and $V_1$ are exceptional,
as they are degenerate representations of $PSU(2,2|4)$.  So we
describe them separately first.  We let $\phi^I$, $I=1,\dots,6$ be
the elementary scalars of ${\cal N}=4$ super Yang-Mills theory.
They transform in the vector representation of the $R$ symmetry
group $SU(4)\cong SO(6)$. Superconformal primaries in $V_0$ and
$V_1$ are the following bilinears in $\phi$:
\eqn\tunbo{\eqalign{V_0:~ &|\psi(0)\rangle \sim
\phi^I\phi^J+\phi^J\phi^I-{1\over 3}\delta^{IJ}\phi^K\phi^K\cr
V_1:~&|\psi(1)\rangle\sim \phi^I\phi^J-\phi^J\phi^I.\cr}} In case
the formula for $|\psi(1)\rangle$ looks strange, note that we need
not worry about bose statistics here, because the fields carry
$U(N)$ gauge indices that are being suppressed. Thus, the
primaries in $V_0$ and $V_1$ have dimension two and non-trivial
$R$-symmetry quantum numbers. By contrast, the primaries
$|\psi(j)\rangle$ in $V_j$, $j\geq 2$, are singlets of the $R$
symmetry, and have dimension $j$.  They can be obtained by
combining two partons in an $SU(4)$ singlet state with relative
angular momentum $j-2$: \eqn\utty{V_j:~|\psi(j)\rangle\sim
\sum_{I=1}^6\sum_{k=0}^{j-2} c_k^{(j-2)}
\partial^k\phi^I\partial^{j-2-k}\phi^I+\dots .}
Here (see for example Eq. 6 in \Mikhailov)
$c_k^{(j-2)}=(-1)^k/k!^2(j-k-2)!^2$.  These precise coefficients
ensure that $V_j$ is a conformal primary.  To make a
superconformal primary, one must add additional terms, also of
angular momentum $j-2$, that are bilinear in fermions or gauge
fields instead of scalars.  We have indicated these terms by the
ellipses in \utty.

We can give as follows a heuristic explanation of why this is the
classification of the $V_j$.  An operator ${\cal O}$ acting on the
vacuum in free field theory creates a two particle state. Consider
the case that particle $A$ is traveling in the $+z$ direction and
particle $B$ in the $-z$ direction.  There are a total of
$16\times 16=256$ helicity states of this type, with 16 for each
particle.  Of the 16 global supercharges (we only consider
supercharges that commute with translations as we have
diagonalized the momentum) of the ${\cal N}=4$ theory, half
annihilate particles moving in the $+z$ direction and a
complementary half annihilate particles moving in the $-z$
direction. So altogether each supercharge acts nontrivially in the
two particle system and it takes $2^{16/2}=256$ states to
represent the 16 supercharges. So in short the global supercharges
act irreducibly on these 256 states, and to classify ${\it N}=4$
multiplets, the relevant variable is the angular momentum, which
is the variable $j-2$ in \utty.  However, this argument breaks
down for small $j$ when the supersymmetries fail to act in a
nondegenerate fashion, and this is how the exceptional
representations $V_0$ and $V_1$ appear.

In section 3.1, an important role is played by the behavior of
$V_j$ under the operator $\sigma$ that exchanges the two spins.
From the above description of the $V_j$, it is clear that $\sigma
V_j =(-1)^j V_j$.  For $j=0,1$, this reflects the fact that the
primary in $V_0$ is symmetric in $I,J$ while that in $V_1$ is
antisymmetric.  For $j\geq 2$, it reflects the fact that a state
of two scalars with relative angular momentum $j-2$ transforms as
$(-1)^{j-2}$ under exchange of the two particles; more explicitly,
one can note that $c^{(j-2)}_k=(-1)^jc^{(j-2)}_{j-2-k}$.

As exploited in \Beisertone, the quadratic Casimir operator $J^2$
of $PSU(2,2|4)$ has the amazing property
\eqn\unfog{J^2V_j=j(j+1)V_j,} rather as if the group were $SU(2)$
instead of $PSU(2,2|4)$, and despite the exceptional nature of
$V_0$ and $V_1$.  Since we need this relation in section 3, we
sketch a proof.  The quadratic Casimir for  $PSU(2,2|4)$ is given
explicitly by \eqn\quacas{\eqalign{J^2 &= \half D^2 + \half
L^\gamma_{\hskip2pt\delta} L^\delta_{\hskip2pt\gamma} + \half \dot
L^{\dot\gamma}_{\hskip2pt{\dot\delta}} \dot
L^{\dot\delta}_{\hskip2pt\dot\gamma} -\half  R^c_d R^d_c -\half
[Q^c_\gamma, S^\gamma_c] -\half [\dot Q^c_\gamma, \dot S^\gamma_c]
-\half \{ P_{\gamma\dot\delta}, K^{\gamma\dot\delta} \}\,.\cr}}
Here $1\le \alpha,\dot\alpha \le 2, 1\le a\le 4$. When acting on
the superconformal primaries $|\psi(j)\rangle$, the Casimir $J^2$
can be rewritten as \eqn\quacastwo{\eqalign{J^2 |\psi(j)\rangle &=
\left(\half D^2 + \half L^\gamma_{\hskip2pt\delta}
L^\delta_{\hskip2pt\gamma} + \half \dot
L^{\dot\gamma}_{\hskip2pt{\dot\delta}} \dot
L^{\dot\delta}_{\hskip2pt\dot\gamma} -\half  R^c_d R^d_c
\right.\cr &\left.+\half \{S^\gamma_c, Q^c_{\gamma}\} +\half
\{\dot S^\gamma_c, \dot Q_{\dot\gamma c} \} -\half [
K^{\gamma\dot\delta}, P_{\gamma\dot\delta}] \right)
|\psi(j)\rangle. \cr}}  We have used the fact that the primary is
annihilated by $S$ and $K$. \quacastwo\ can be easily evaluated
using the commutation relations as \eqn\quath{\eqalign{J^2
|\psi(j)\rangle &= (\half D^2 + s_1(s_1 + 1) + s_2 (s_2 + 1)
-\half R^c_d R^d_c + 2 D) |\psi(j)\rangle\,.\cr}} Here $s_1$ and
$s_2$ are the spin quantum numbers of an operator; a general
operator transforms under the four-dimensional rotation group
$SO(4)\cong SU(2)\times SU(2)$ with spins $(s_1,s_2)$. For $j\ge
2$, $|\psi(j)\rangle$is annihilated by the $R^c{}_d$ as it is an
$SU(4)_R$ singlet, and has dimension $j$ and $s_1=s_2=(j-2)/2$. So
we get \eqn\quafiv{\eqalign{J^2 |\psi(j)\rangle &= \left(\half j^2
+ \left({j\over 2} - 1\right) {j\over 2} 2 + 2j\right)
|\psi(j)\rangle = j(j+1) |\psi(j)\rangle\,.\cr}}
For the cases $j=0$ and $j=1$, the conformal dimension is two, and the
spins $s_1$ and $s_2$ are zero. The $SU(4)_R$ contribution is
$-\half R^c_d R^d_c = -6 $ for the traceless symmetric tensor 
(the $j=0$ case) giving $J^2 |\psi(0)\rangle = 0$,
and $-\half R^c_d R^d_c = - 4 $ for the adjoint representation
(the $j=1$ case) giving $J^2 |\psi(1)\rangle = 2$.

\vskip40pt

\noindent{\bf Acknowledgements:}

We thank P. Svrcek for discussions.

CRN was partially supported by the NSF Grants PHY-0140311 and
PHY-0243680.
Research of EW was supported in part by
NSF Grant PHY-0070928.
LD thanks Princeton University for its hospitality,
and was partially supported by the U.S. Department of Energy,
Grant No. DE-FG02-03ER41262.

Any opinions, findings, and conclusions or recommendations
expressed in this material are those of the authors and do not necessarily
reflect the views of the National Science Foundation.

\listrefs

\end